\documentclass[a4paper,12pt]{cernart} 
\usepackage[dvips]{graphicx}
\begin{document}
\newcommand{\Lt}{\lambda_\theta}
\newcommand{\Lp}{\lambda_\phi}
\newcommand{\Ltp}{\lambda_{\theta \phi}}
\newcommand{\qqbar}{\mathrm {q}\overline \mathrm {q}}
\newcommand{\dNdO}{\mathrm{d}N\!/\!\mathrm{d}\Omega}
\begin{titlepage}
\docnum{CERN--PH--EP--2010--080}
\date{7 December  2010}
\vspace{1.2cm}
\title{ANGULAR DISTRIBUTION AND ROTATIONS OF FRAME\\ IN VECTOR MESON DECAYS INTO LEPTON PAIRS}
\begin{Authlist}
Sandro Palestini
\Instfoot{a1}{CERN, 1211 Geneva 23, Switzerland}
\end{Authlist}
\vspace{2.5cm}
\begin{abstract}\noindent
We discuss how the angular distribution of lepton pairs 
from decays of vector mesons depends on the choice of 
reference frame, and provide a geometrical description of the transformations
of the coefficients of the angular distribution. Invariant expressions involving all coefficients 
are discussed, together with bounds and consistency relations.
\end{abstract}
\vspace{3cm}
\end{titlepage}
\section{Introduction}
%
The study of the angular distribution of lepton pairs in hadron collisions 
allowed the test of the Drell-Yan model \cite{ref:DY,ref:CS}
and of its corrections in perturbative QCD \cite{ref:LT},
and verified the electro-weak couplings of W and Z bosons.  
More recently, angular distributions have been suggested in order to 
discriminate among different production models for
$\qqbar$ states \cite{ref:review, ref:NRQCD, ref:singlet} in hadron collisions.

Available results on $\mathrm {c}\overline \mathrm{c}$ and $\mathrm {b}\overline \mathrm{b}$ 
resonances \cite{ref:psi,ref:Y-colliders,ref:Y-ft,ref:phenix} do not 
provide a complete and consistent picture \cite{ref:review, ref:Faccioli-lungo}.
This is in part due to limited acceptance ranges, which
prevent a full analysis of the polarization of the resonance, and also to  
different choices of reference system. New data from LHC is eagerly awaited. 

Recent studies \cite{ref:Faccioli-lungo,ref:Faccioli}  have discussed the relevance of taking 
directly into account properties related to the description of the process 
in different frames, and have specified rotation-invariant quantities that 
provide information on intrinsic, frame independent properties of the 
polarization of the $\qqbar$ state.  

The work presented here extends along the same direction, under the 
assumption of parity conservation. 
We provide a geometrical description of the transformation of the 
different terms of the angular distribution, and specify new invariant 
quantities that relate all the components.
We also establish physical bounds and consistency relations among all terms 
of the angular distribution. 

\section{Angular distributions and rotations of frame}
The angular distribution of the lepton pair in its rest frame is described by:
\begin{equation} \label{eq:basic}
\dNdO = [1 + \Lt \cos ^2\!\theta + \Lp \sin^2\!\theta \cos(2\phi) + 
\Ltp \sin(2\theta) \cos\phi]/(1+\Lt /3) \; .
\end{equation}
Here $\theta$ and $\phi$ identify as usual the direction of the positive lepton, with $\phi$=0
corresponding to the production plane of the vector meson.  In hadron collisions, different choices have been made for the axis $z$, including those that approximate the collision axis, like the 
Collins-Soper (CS) frame \cite{ref:CS} and the {\em t}-channel/Gottfried-Jackson 
frame \cite{ref:GJ},  
or the helicity (H) frame, that selects the direction of the pair in the 
c.m.\ system of the collision. 
Equation~(\ref{eq:basic}) makes the assumption of parity conservation and 
symmetry for reflection on the production plane, and its validity covers both the case of elementary processes, where the coefficients $\Lt$, $\Lp$ and $\Ltp$ are directly related to the helicity amplitudes of the vector meson,  or the case of different, uncorrelated processes contributing to the angular distribution.

A change of frame ({\em e.g.\/}: between CS and H) corresponds to a rotation 
about the $y$-axis. Under a rotation by $\delta$ 
the functional dependences in (\ref{eq:basic}) are preserved 
if the coefficients of the angular distribution are transformed as:
\begin{eqnarray} 
1 \;\Rightarrow 
      &1+\frac{1}{2} \sin ^2 \! \delta \cdot (\Lt-\Lp)+\frac{1}{2} \sin (2\delta) \cdot  \Ltp  
             & =\; N  \nonumber \\ 
\Lt \; \Rightarrow
      &\left(1-\frac{3}{2} \sin ^2 \!\delta \right)\cdot \Lt +\frac{3}{2} \sin^2 \! \delta \cdot \Lp 
- \frac{3}{2} \sin (2\delta) \cdot \Ltp 
             & = \; N\, \Lt^\prime \nonumber \\ 
\Lp \; \Rightarrow
       &\left(1-\frac{1}{2} \sin^2 \!\delta \right) \cdot \Lp + \frac{1}{2} \sin^2 \!\delta \cdot \Lt
+ \frac{1}{2} \sin (2\delta) \cdot \Ltp 
             & = \; N\, \Lp^\prime \nonumber \\ 
\Ltp \; \Rightarrow
        &\cos (2\delta) \cdot \Ltp + \frac{1}{2} \sin (2\delta)\cdot  (\Lt-\Lp) 
              & = \; N\, \Ltp ^\prime \; , \label{eq:transf}
\end{eqnarray}
where $N$ enters as a change in the normalization.  

From (\ref{eq:transf}) we see that we can always define a rotation to a 
frame in which $\Ltp=0$, by means of a rotation by the angle $\delta^o$ 
(or $\delta^o+\pi$) satisfying:
\[ 
\tan(2\delta^o)= -2\,\Ltp/(\Lt-\Lp)\; .
\] 

\section{Properties in the oriented frame}
If $\Ltp$=0 the reference frame is {\em oriented} 
along the principal axes of symmetry of the angular distribution. 
We shall first discuss properties of the components of the angular 
distribution for this choice of frame.
Particular cases of angular distribution include: (a)~{\em longitudinal\/} polarization 
along the $z$-axis ($\Lt$=--1, $\Lp$=$\Ltp$=0):
\[ 
\dNdO\,_{\rm Long-z} = {\textstyle \frac{3}{2}}(1-\cos ^2\!\theta) 
                     = {\textstyle \frac{3}{2}}(1-\hat{z}^2)
                     = {\textstyle \frac{3}{2}}(\hat{x}^2+\hat{y}^2) \: ;
\] 
(b)~{\em transverse(-z)} polarization, 
which conventionally refers to the case $\Lt$=+1, $\Lp$=$\Ltp$=0 and is described by:
\[ 
\dNdO\,_{\rm Trans-z} = {\textstyle \frac{3}{4}}(1+\cos ^2\!\theta) 
                      = {\textstyle \frac{3}{4}}(1+\hat{z}^2)\; .
\] 
Transverse polarization implies that the contributing processes select the vector meson in 
a state of definite helicity  ($J_z$=+1 or $J_z$=--1) along the same axis $z$. 
Neglecting the mass of the leptons, the pairs have always $J_{z^\prime}$=$\pm $1
along their axis, and the $1+\cos^2\!\theta$ distribution is determined 
by the sum of the squares of the d$^1_{+1+1}$ and d$^1_{-1+1}$ terms 
in the Wigner matrix. Similarly, the longitudinal polarization corresponds to $J_z$=0 along 
the same axis for all contributing processes. 

The transverse(-$z$) angular distribution can be expressed as 
a sum of longitudinal--$x$ and longitudinal--$y$ distributions:
\[ 
\dNdO\,_{\rm Trans-z} = {\textstyle \frac{3}{4}}(1+\hat{z}^2)
   ={\textstyle \frac{3}{4}}(2-\hat{x}^2-\hat{y}^2)
   ={\textstyle \frac{1}{2}}\,\dNdO\,_{\rm Long-x}
     +{\textstyle \frac{1}{2}}\,\dNdO\,_{\rm Long-y}\: .
\] 

A similar relation can be written for any angular distribution described in the oriented frame. In fact, 
terms proportional to $\hat x\hat y$, $\hat z\hat y$ and $\hat z\hat x$
cannot be present because of the symmetry assumptions contained in (\ref{eq:basic}) and in the definition of oriented frame. We therefore argue that 
any angular distribution in the oriented frame can be described 
by the sum of the three (longitudinal) contributions referred to the reference axes:
\begin{equation} \label{eq:abc}
\dNdO(a,b,c) = a\cdot \dNdO\,_{\rm Long-z} 
                            +b\cdot \dNdO\,_{\rm Long-x} 
                            +c\cdot \dNdO\,_{\rm Long-y} \; .
\end{equation}
The coefficients $a$, $b$ and $c$ are non--negative and satisfy 
$a+b+c=1$, so that we can choose $b$, $c\,$ as independent variables and write:
\begin{equation} \label{eq:lim-abc}
0\leq b \leq 1\;, \;\;\; \;  0\leq c \leq 1-b \; .
\end{equation}
Writing (\ref{eq:abc}) explicitly and comparing with (\ref{eq:basic}), 
we obtain the relations: 
\begin{eqnarray*}
\Lt=&\;\: {\displaystyle \frac {1-3\,a}{1+a}}&=\quad \frac{3(b+c)-2}{2-b-c} \; , \\
\Lp=&{\displaystyle {}- \frac{b-c}{1+a}} &= {}-\cos(2\chi)\frac{1+\Lt}{2} \; ,
\end{eqnarray*}
where $\chi$=$\arctan\sqrt{c/b}$. From these equations and the limits 
(\ref{eq:lim-abc}) we see that:\footnote{
The limit on $\Lp$ could have also been obtained from (\ref{eq:transf}) and 
requiring that $|\Lt^\prime|\leq$1 for any angle $\delta$ \cite{ref:Faccioli}.}
\begin{equation} |\Lt |\leq 1  \; , \qquad 
|\Lp |\leq {\textstyle \frac{1}{2}}(1+\Lt)\; .  \label{eq:limit2}
\end{equation}

\section{Frame rotations in the space of the coefficients}
\begin{figure}
\begin{center}
\includegraphics*[scale=0.65]{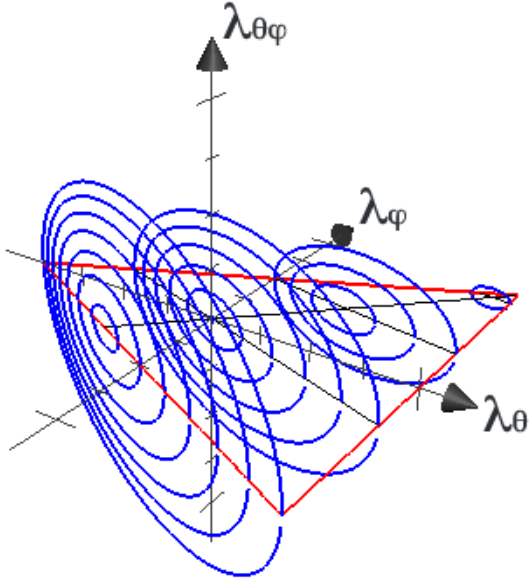}
\caption {Ellipses of rotation in the space of the coefficients of the angular
distribution. Four sets, corresponding to $\tilde\lambda=$ -1, 0, 3, 37 
(see Section~\ref{sec:invar}) are drawn.}
\label{fig:ellipses}
\end{center}
\end{figure}

Applying the transformation~(\ref{eq:transf}) to sets of values on an oriented frame ($\Ltp$=0),
we can analyze properties of the angular distribution in any frame. Figure~\ref{fig:ellipses} 
illustrates the transformation in the three-dimensional space of $(\Lt,\Lp,\Ltp)$: the coefficients vary following closed loops, with mirror symmetry on the plane $\Ltp=0.$

\begin{figure}
\begin{center}
\includegraphics*[scale=0.65]{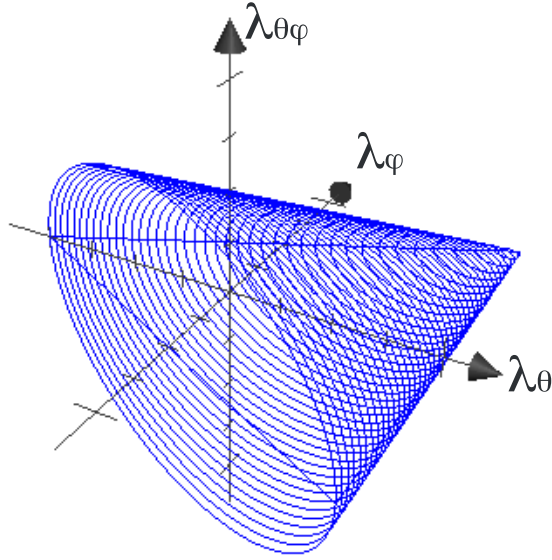}
\caption{Envelope of the ellipses of maximum size.}
\label{fig:cone}
\end{center}
\end{figure}

Before proving that the loops are ellipses (Section~\ref{sec:limit3}), we can list the following properties: 
(a)~the loops are wound about the line: $\Lt$ =$\Lp$\,, $\Ltp$=0; 
(b)~they are contained in planes orthogonal to the plane $\Ltp$=0;
(c)~for each plane, the ellipsis of largest 
size is the one connecting the boundary lines $\Lt$=1 and $\Lp$=+0.5($\Lt$+1) 
on the plane $\Ltp$=0.
Figure~\ref{fig:cone} shows the envelope of the ellipses of maximum 
size, which is a cone in the 3D space. Clearly, the limits (\ref{eq:limit2}) are valid in any frame and the range of $\Ltp$ depends on $\Lt$ and $\Lp$. 

The geometrical description illustrates properties of the dependence of the coefficients $\Lt$, 
$\Lp$, $\Ltp$ on the rotation angle $\delta$ that may not be immediately 
evident from the the transformations~(\ref{eq:transf}), including:
(a) the maximum (minimum) of $\Lt$ coincides with the minimum (maximum) of $\Lp$; 
(b) the points above correspond to the zeroes of $\Ltp$; 
(c) the mid--range value of $\Lt$ coincide with that of $\Lp$;
(d) the extremes $\pm |\Ltp|$ coincide with the mid--range values in $\Lt$, $\Lp$.  


\section{Invariant quantities} \label{sec:invar}
Intrinsic, frame--independent characteristics of the angular distribution 
are those linked to properties of the ellipses, and not depending on the
position of the point as it moves along the loops when the reference frame is 
rotated about the $y$--axis.
An example of invariant is the specification of the plane that contains 
a set of ellipses, which we can write as:
\begin{equation}   \Lp{}-1\:=\: {}-{}(\Lt+3)/(3+\tilde\lambda) \; , 
\label{eq:L_tilde} 
\end{equation}
where $\tilde\lambda$
is the invariant quantity discussed in 
reference \cite{ref:Faccioli}, which we now interpret as the projection of 
the ellipses onto the plane $\Ltp=0$. The invariance of (\ref{eq:L_tilde})
means that it remains valid, for the same value of $\tilde\lambda$, if 
the coefficients are transformed as in (\ref{eq:transf}). In other words, 
$\tilde\lambda=(\Lt +3\Lp )/(1-\Lp)$ is an invariant expression.
  
Other expressions invariant for changes of frame  
may be associated to the size of the axes, or to the 
ranges in $\Lt$, $\Lp$ and $\Ltp$. These invariants will involve all three 
coefficients of the angular distribution.  
An example is given by the differences $\Delta_\pm$=($\Lt{}^e$ -$\Lp{}^e)_\pm$, defined at 
the extremes of the ellipsis where $\Ltp{}^e=0$, with $\pm$ specifying 
the side with positive or negative value.  Given any set of values 
$(\Lt,\Lp,\Ltp)$, with $\Delta$=$\Lt$-$\Lp$, we find:
\begin{equation} \label{eq:Delta+-} 
\Delta_\pm=\frac{\pm\left(\Delta^2\,+\,4\,\Ltp{}^2\right)}
{\left(1+\Delta/4\right)\sqrt{\Delta^2\,+\,4\,\Ltp{}^2}
\mp \left(\Delta^2\,+\,4\,\Ltp{}^2\right)/4} \; .
\end{equation}
The invariance of (\ref{eq:Delta+-}) corresponds to the invariance of the 
simpler expression:
\[ 
\lambda^*\equiv \frac{1+\Delta/4}
{\sqrt{\Delta^2\,+\,4\,\Ltp{}^2}}=
\frac{1\,+\,(\Lt-\Lp)/4}{\sqrt{(\Lt-\Lp)^2\,+\,4\,\Ltp{}^2}} \;.
\] 

As an additional example, the invariant corresponding to the maximum range 
$\pm|\Ltp|^{\mathrm max}$ is related to $\Delta_\pm$ through:
\[ 
|\Ltp|^{\mathrm max}=\frac{\Delta_+}{2\sqrt{1+\Delta_+/2}}
                    =\frac{-\,\Delta_-}{2\sqrt{1+\Delta_-/2}}\;.
\] 

We should keep in mind that a single invariant
does not fully specify the intrinsic properties of the  angular 
distribution. 
For example, all the ellipses contained in the same vertical planes in 
Figure~\ref{fig:ellipses} share the value of $\tilde\lambda$. 
In other words, sets $(\Lt,\Lp,\Ltp)$ 
providing the same value of $\tilde\lambda$ are not necessarily linked 
by a rotation of frame about the {\em y}--axis 
--- while the reverse statement is true. 
Similarly, the value of $\lambda^*$ is shared by 
ellipses with the same value of $\Lt$-$\Lp$ on the plane $\Ltp$=0.
On the other hand, the combination of these two invariants identify entirely 
the  intrinsic angular distribution.\footnote
{As discussed in Refs.~\cite{ref:Faccioli-lungo} and \cite{ref:Faccioli} for $\tilde\lambda$, 
the invariants may be interpreted in terms of properties of 
individual processes contributing to the angular distribution, defined in 
specific, process dependent frames. For $\lambda^*$, information on 
$\Lt$-$\Lp$ in the oriented frame is provided. The term $\Lp$
describes asymmetries between $x$ and $y$. For example, for $\Lt$=1, 
$\Lp$ specifies the orientation of a polarization orthogonal to the $z$ axis,
with longitudinal--$x$ (--$y$)  for $\Lp$=${}-1$ (+1), and with circular 
polarizations: $(x\pm{\mathrm i}y)/\sqrt{2}$ for $\Lp$=0.}


\section{Bounds on the coefficients} \label{sec:limit3}
Figure~\ref{fig:cone} shows that the largest allowed values of $\Ltp$ occurs 
for $\Lt$=0, $\Lp$=--0.5, where $|\Ltp|^{\mathrm max}$=1/$\sqrt{2}$, while it 
vanishes on the opposite sides of the allowed triangle: $\Lt$=1, 
$\Lp$=0.5(1+$\Lt$). 

\begin{figure}
\begin{center}
\includegraphics*[scale=0.59]{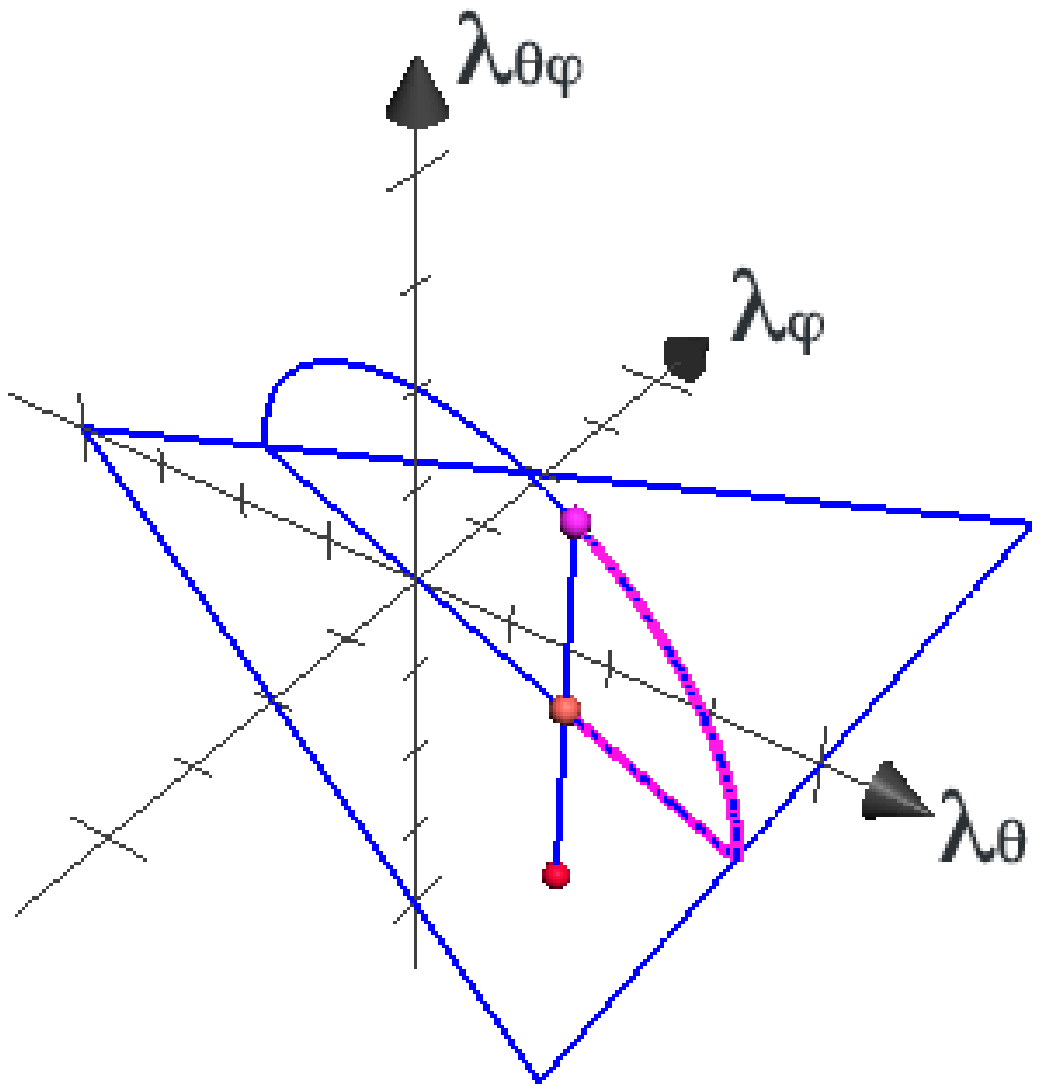}
\caption{Illustration of the maximum allowed values of $\Ltp$ for given values 
of $\Lt$ and $\Lp$.}
\label{fig:max}
\end{center}
\end{figure}

We can obtain the maximum allowed value $|\Ltp|^{\mathrm max}$ compatible with 
given values of $\Lt$ and $\Lp$ with the following procedure: 
(a)~define with (\ref{eq:L_tilde}) the line corresponding to $\Lt$, $\Lp$ 
and consider the segment reaching the sides of the triangle defined by (\ref{eq:limit2}); 
(b)~use the transformation~(\ref{eq:transf}) and compute the rotation angle that maps  
one end-point of the segment into a point with the given value $\Lt$ (or $\Lp$); 
(c)~using the same angle, $|\Ltp|^{\mathrm max}$ is obtained through the corresponding rotation from the end-point.
The procedure is illustrated in Figure~\ref{fig:max}. 
The result is the equation:
\begin{equation} \label{eq:Ltp-limit}
|\Ltp|^{\mathrm max}=\frac{1}{2}\:\sqrt{(1-\Lt)(\Lt+1-2\Lp)}\quad.
\end{equation}
Together with (\ref{eq:limit2}), this expression completes the limits and the consistency relations among the coefficients $\Lt$, $\Lp$ and $\Ltp$.\footnote
{The bound to $\Ltp$ may be compared to the less constraining relation: $|\Ltp |\!\leq$ 0.5(1-$\Lp$) for $\Lp\!\geq$--1/3, 
$|\Ltp|\! \leq\! \sqrt {-2\Lp(1+\Lp)}$ for $\Lp\!\! \leq$--1/3, from Ref.~\cite{ref:Faccioli-lungo}.  These inequalities provide the envelope to the 
projection of (\ref{eq:Ltp-limit}) on the $\Lp$ axis, {\em i.e.\/}: they are numerically equal if $\Lt$ is chosen to maximize  (\ref{eq:Ltp-limit}) locally.} 
The allowed range of values is illustrated in Figure~\ref{fig:cone}.

Furthermore, equation (\ref{eq:Ltp-limit}) specifies the geometry of the transformation of the 
coefficients. Under a rotation of frame, the variations of $\Lt$ and $\Lp$ are constrained by 
(\ref{eq:L_tilde}). If we substitute $\Lp$ retaining $\Lt$ as independent variable, the equation for 
$|\Ltp|^{\mathrm max}$  can be written as:
\[ 
|\Ltp|^{\mathrm max}=\frac{1}{2}\:
  \sqrt{\frac{\tilde\lambda+5}{\tilde\lambda+3}}\:
  \sqrt{(1-\Lt)\left(\Lt-\frac{\tilde\lambda-3}{\tilde\lambda+5}\right)}\quad.
\]
This equation describes an ellipsis in the $\Lt \Ltp$ plane. Since
$\Lt$ and $\Lp$ are linearly related, a corresponding equation describes the 
loops in Figure~\ref{fig:cone}. 
A similar relation can be written for loops not reaching the maximum 
extension in $\Lt$ or $\Lp$, proving that any transformation~(\ref{eq:transf})
describes an ellipsis in the 3D space of the coefficients.
\section{Conclusions}
Understanding the implications of the choice of reference frame 
is relevant for the study of of $\qqbar$ resonances decaying to lepton pairs, 
and we have provided a geometrical interpretation of the transformations of the
coefficients of the angular distribution.   
For transverse momentum of the pair comparable or larger than its invariant mass, 
the rotation angle between the CS and the H frames may be comparable to   
$\pi$/2 over large ranges of rapidity, with corresponding large 
{\em rotations} in the $\Lt\Lp\Ltp$ space. Indeed points close to the boundary 
$\Lt$=1 ($\Ltp$=0) may be mapped into those close to $\Lp$=1/2(1+$\Lt$).
On the other hand, points near the vertex of the cone ($\Lt$=$\Lp$=1,
$\Ltp$=0) or to the line $\Lt$=$\Lp$, $\Ltp$=0 are scarcely changed by 
a different choice of reference frame.

Besides the geometrical description, our analysis has provided further 
insight on the subject of invariant quantities, and has specified in detail
the range of allowed values for the coefficient $\Ltp$.

We thank Vato Kartvelishvili for calling our attention to this topic and for helpful discussions.

\end{document}